\documentclass[%
 reprint,
superscriptaddress,
 amsmath,amssymb,
 aps,
 prl,
]{revtex4-1}

\usepackage{dcolumn}
\usepackage{float}
\usepackage{bm}
\usepackage{comment} 
\usepackage{hyperref}
\usepackage{titlesec}
\usepackage[titletoc,toc,title]{appendix}
\usepackage{amssymb,amsmath,graphicx,array,color,amsthm}
\usepackage{nccmath}
\usepackage{caption}
\usepackage[skip=-5pt]{subcaption}
\usepackage{amscd}
\usepackage{multirow}

\def \BE {\begin{equation}}
\def \EE {\end{equation}}
\def \BEA {\begin{eqnarray}}
\def \EEA {\end{eqnarray}}

\def \rmd {\,\mathrm{d}\,}

\begin{document}
\preprint{AIP/123-QED}

\title{Direct and inverse cascades in turbulent Bose-Einstein condensate}
\author{Ying Zhu}
\email{yzhu@unice.fr}
\affiliation{Universit\'{e} C\^{o}te d'Azur, CNRS, Institut de Physique de Nice (INPHYNI), Parc Valrose, 06108 Nice, France}
\author{Boris Semisalov}
\affiliation{Universit\'{e} C\^{o}te d'Azur, Observatoire de la C\^{o}te d'Azur, CNRS, Laboratoire Lagrange, Boulevard de l'Observatoire CS 34229 -- F 06304 Nice Cedex 4, France}
\affiliation{Novosibirsk State University, 1 Pirogova street, 630090 Novosibirsk, Russia}
\affiliation{Sobolev Institute of Mathematics SB RAS, 4 Academician Koptyug Avenue, 630090 Novosibirsk, Russia}
\author{Giorgio Krstulovic}
\affiliation{Universit\'{e} C\^{o}te d'Azur, Observatoire de la C\^{o}te d'Azur, CNRS, Laboratoire Lagrange, Boulevard de l'Observatoire CS 34229 -- F 06304 Nice Cedex 4, France}
\author{Sergey Nazarenko}
\affiliation{Universit\'{e} C\^{o}te d'Azur, CNRS, Institut de Physique de Nice (INPHYNI), Parc Valrose, 06108 Nice, France}


\begin{abstract} 
  When a Bose-Einstein condensate (BEC) is driven out of equilibrium, density waves interact non-linearly and trigger  turbulent cascades. In a turbulent BEC, energy is transferred towards small scales by a direct cascade, whereas the number of particles displays an inverse cascade toward large scales. In this work, we study analytically and numerically the direct and inverse cascades in wave-turbulent BECs. We analytically derive the Kolmogorov-Zakharov spectra, 
  including the log-correction to the direct cascade scaling and the universal pre-factor constants for both cascades.
  We test and corroborate our predictions using high-resolution numerical simulations of the forced-dissipated Gross-Pitaevskii model in a periodic box and the corresponding wave-kinetic equation. Theoretical predictions and data are in excellent agreement, without adjustable parameters. Moreover, in order to connect with experiments, we  test and  validate our theoretical predictions using the Gross-Pitaevskii model with a confining cubic trap. Our results explain previous experimental observations and suggest new settings for future studies.
\end{abstract}

\maketitle  

In many nonlinear systems, a non-trivial out-of-equilibrium state emerges when dissipation and injection of some invariant (typically energy) occur at very different scales. Such states are {often} characterized by a constant flux across scales of the invariant in a cascade process. In general terms, the process where an invariant is transferred from large to small scales is called a \emph{direct cascade}, whereas the opposite -- an \emph{inverse cascade}. 
Such cascades play a central role in hydrodynamic and wave turbulence. In the former case, they are powered by hydrodynamic vortex interactions, whereas in the latter case -- by interactions of random  waves. There are numerous important physical examples of wave turbulence (WT) in Nature: among many others, turbulence of inertial and internal waves in rotating stratified fluids \cite{Caillol_KineticEquationsStationary_2000,Galtier_WeakInertialwaveTurbulence_2003}, gravitational waves \cite{galtier2017turbulence}, Kelvin waves in superfluids vortices~\cite{Lvov_WeakTurbulenceKelvin_2010} and Bose-Einstein condensates (BECs) \cite{dyachenko1992optical}. Unlike hydrodynamic turbulence -- where most predictions remain phenomenological -- when waves are ``weak'', WT theory furnishes analytical predictions for the wave excitation spectrum which can be found as  exact solutions of an associated wave kinetic equation (WKE). It expresses the spectrum in terms of the flux and wave numbers, predicts the direction of the cascades and provides the values of the universal proportionality constants~\cite{nazarenko2011wave}.

Remarkably, recent experiments with BECs have succeeded in achieving controlled WT processes in the direct energy cascade setting~\cite{navon2016emergence,navon2019synthetic}. Intriguingly, those experiments measured a notably steeper (in wave number) spectrum than the one predicted by the theory~\cite{dyachenko1992optical}. In addition to being a fundamentally important state of matter, BECs have great potential as a platform for experiments in turbulence, both vortex and wave kinds. This richness is due to the close analogy between the BEC motion and the classical fluid flow. Because of the versatility of current optical techniques, BEC experiments allow a great deal of flexibility often unavailable in classical fluid experiments. Moreover, the Gross-Pitaevskii equation (GPE), which describes BEC dynamics, is a universal nonlinear model whose importance spans diverse physical systems, particularly in optics, plasmas and water wave theory~\cite{dyachenko1992optical}.

In this Letter, we study the direct and inverse cascades of turbulent BECs. We use the GPE description of a BEC and its associated WKE. We obtain new analytical predictions, which explain the steeper spectrum observed in \cite{navon2016emergence}, and provide the values of the dimensionless universal constants in the direct energy and inverse particle cascades. Our predictions are then tested by high-resolution direct numerical simulations of the GPE and its associated four-wave WKE.

The dimensionless GPE equation for the complex wave function $\psi({\bf x},t)$ is 
 \begin{equation}
   \frac{\partial\psi ({\bf x},t)  }{\partial t}=i \left [ \nabla^{2}  -\left|\psi({\bf x},t) \right|^{2} +U({\bf x}) \right ]\psi({\bf x},t) \,,
  \label{GPE}
 \end{equation} 
 where $U({\bf x})$ is an external trapping potential. Eq.~\eqref{GPE} is obtained from the standard dimensional GP by a proper rescaling of time and space; see Supplemental Material (SM). For simplicity, in the first part of this Letter, we study a homogeneous ($U=0$) three-dimensional BEC. We consider the GPE in a triply-periodic cube of side $L$ and volume $V=L^3$. 
 GPE \eqref{GPE} conserves the total number of particles and energy per unit of volume
 \begin{eqnarray}
N &=&\frac{1}{V}\int_V |\psi(\mathbf{x}, t)|^2 \rmd\mathbf{x} ,\label{eq:cons-N} \\
H&=&\frac{1}{V}\int_V \left[ |\nabla \psi(\mathbf{x}, t)|^2 + \frac{1}{2}|\psi(\mathbf{x}, t)|^4 \right] \rmd\mathbf{x} \, \label{eq:cons-H},
\end{eqnarray}
respectively.

When the condensate is negligible, the WT theory for the GPE formulates an asymptotic closure for the waveaction spectrum $n_{\bf k}(t)\equiv n({\bf k},t) = \frac{V}{(2\pi)^3}\langle |\hat \psi_{\bf k}(t)|^2 \rangle,$
where $\hat \psi_{\bf k}(t)$ is the Fourier transform of
$\psi(\mathbf{x}, t)$, and the brackets denote averaging over the initial wave statistics.
The WT closure is derived under assumptions of small nonlinearity and random initial phases and amplitudes of waves~\cite{ZLF,nazarenko2011wave}. It furnishes a wave-kinetic equation (WKE) with four-wave interactions \cite{ZAKHAROV1985285,dyachenko1992optical}. For an isotropic spectrum, which depends only on the magnitude of the wave vector $k=|{\bf k}|$, it is given by
\cite{semikoz1995kinetics,zhu2022testing}
  \begin{eqnarray}\label{WKE}
    &\frac{\partial n_{k}}{\partial t}= \text{St}_k(t)\equiv
    \tfrac{32\pi^3}{k}\medint\int
    \limits_{k_i>0}
    \min\left(k,k_1,k_2,k_3\right)k_1k_2k_3  \quad\,  \\
    & n_{k}  n_{{k_1}}n_{{k_2}}n_{{k_3}} \left(\tfrac{1}{n_{k}} + \tfrac{1}{n_{{k_1}}} - \tfrac{1}{n_{{k_2}}}-\tfrac{1}{n_{{k_3}}}\right) \delta(\omega^{01}_{23}) \mathrm{d}k_1\mathrm{d}k_2\mathrm{d}k_3\,,  \nonumber
    \end{eqnarray}  
where $\omega^{01}_{23} \equiv \omega_{k} + \omega_{k_1} -\omega_{k_2} -\omega_{k_3}$
with $\omega_k = \omega(k)$ being the frequency given by the dispersion relation $\omega_k =k^2$.
 
The WKE conserves the densities of the number of particles and the energy,
\begin{equation}\label{eq:cons-WKE}
N =  4\pi \medint \int_0^\infty  k^2 \, n_{k} \, \mathrm{d}k \,,\quad
E =  4 \pi\medint \int_0^\infty   k^4 \, n_{k} \,\mathrm{d}k\,, 
\end{equation}
which coincide with \eqref{eq:cons-N} and with the first term of the integrand in \eqref{eq:cons-H} (the second term is small), respectively.

It is well known that the four-wave WKE may have two Kolmogorov-Zakharov (KZ) type
non-equilibrium stationary solutions. 
KZ solutions are expected in forced-dissipated wave systems in which WT is forced and dissipated at small and large wave-vectors respectively for a direct cascade, and vice versa for an inverse cascade.
 
 To find stationary solutions, we assume a power law spectrum 
 in the form
 $n_k=Ak^{-2x}$. The right-hand side (RHS) of \eqref{WKE}  becomes $\text{St}_k=4\pi^3A^3k^{4-6x}I(x)$, where
 \begin{eqnarray}\label{eq:Ix}
&I(x)=\medint\int \left[\min\left(1,q_1,q_2,q_3 \right)\right]^{1/2} \left(1+q_1^x-q_2^x-q_3^x\right)\qquad\\
&\left( q_1q_2q_3 \right)^{-x} 
\delta \left(q^{01}_{23}\right)
\mathrm{d}q_1\mathrm{d}q_2\mathrm{d}q_3\,,\,\,\,  q_i>0 \nonumber\,,
 \end{eqnarray}  
is the dimensionless collision term depending only on $x$,
and $\delta \left(q^{01}_{23}\right)=\delta\left(1+q_1-q_2-q_3\right)$. 
Zakharov's transformation (ZT) allows finding stationary solutions (zeros of $I(x)$) by mapping the integration subdomains into a single triangle \cite{nazarenko2011wave} as follows
\begin{equation}\label{ZT}
\begin{split}
    & q_2=\tfrac{1}{\tilde{q_2}}\,, q_1=\tfrac{\tilde{q_3}}{\tilde{q_2}}\,,  q_3=\tfrac{\tilde{q_1}}{\tilde{q_2}}\,,\,  \text{for}\,\, q_2>1\,,0<q_3<1\,,\\ 
    & q_3=\tfrac{1}{\tilde{q_3}}\,,  q_1=\tfrac{\tilde{q_2}}{\tilde{q_3}}\,,  q_2=\tfrac{\tilde{q_1}}{\tilde{q_3}}\,,\, \text{for}\,\, 0<q_2<1\,,q_3>1\,, \\
    & q_1=\tfrac{1}{\tilde{q_1}}\,,  q_2=\tfrac{\tilde{q_3}}{\tilde{q_1}}\,,  q_3=\tfrac{\tilde{q_2}}{\tilde{q_1}}\,,\,  \text{for}\,\,  q_2\,,q_3>1\,. 
\end{split}
\end{equation}
After dropping tildes, $I(x)$ becomes
\begin{eqnarray}\label{eq:Iztx}
&I_{\rm ZT}(x)=\medint\int 
q_1^{1/2-x} 
 \left(q_2q_3 \right)^{-x}
\left(1+q_1^x-q_2^x-q_3^x\right) \quad \\
&\left(1+q_1^y-q_2^y-q_3^y\right)
\delta \left(q^{01}_{23}\right)
\mathrm{d}q_1\mathrm{d}q_2\mathrm{d}q_3\,,\quad 
0<q_i<1\,,\nonumber
\end{eqnarray} 
where $y=3x-7/2$.
$I_{\text{ZT}}(x)=0$ has two apparent solutions $x=7/6$ and $x=3/2$, corresponding to the non-equilibrium stationary inverse cascade $n_k\sim k^{-7/3}$ and direct cascade $n_k\sim k^{-3}$ respectively. 
One should always substitute these candidate $x$ values into $I(x)$ to ensure that the resulting integral is convergent and equal to zero, since
 ZT is not an identity transformation.
 Such an integral convergence, called the interaction locality, physically means that wave quartets with similar values of  wave numbers dominate the nonlinear evolution. 
Mathematically, violation of locality simply means that the considered spectrum is not a valid stationary solution of the WKE.  
Note that under the locality assumption $I_{\text{ZT}}(x)=I(x)$.

Consider first the inverse cascade of particles.  According to (\ref{WKE}, \ref{eq:cons-WKE}), the spectral flux of particles through the sphere of radius $|k|$ on the spectrum $n_k=Ak^{-2x}$ is
\begin{equation} \label{eq:Q_k}
    Q(k)\equiv-4\pi \medint \int_0^k  \kappa^2 \, \text{St}_{\kappa} \, \mathrm{d}\kappa=8\pi^4 A^3k^{7-6x}\tfrac{I_{\rm ZT}(x)}{3x-7/2}\,.
\end{equation}
When $x \to 7/6$, one can use the L’Hopital rule  to derive a constant ($k$-independent) particle flux  thanks to the locality of $I(7/6)$ (see SM): $Q_0=8\pi^4A^3I'_{\text{ZT}}(7/6)/3$,  where $I'_{\text{ZT}}(x)=\rmd I_{\text{ZT}}(x)/\rmd x$. 
Thus, $A=C_{\rm i} |Q_0|^{1/3}$ where $C_{\rm i} >0$ is a dimensionless universal  constant  (recall that $Q_0<0$, but $I_{\text{ZT}}'(7/6)<0$ in SM).
We calculate  $C_{\rm i}$ analytically in SM, and write the resulting  KZ spectrum for the inverse cascade as 
 \begin{equation} \label{eq:ic}
  n_{ k} =C_{\rm i} |Q_0|^{1/3} k^{-7/3}\,,\quad C_{\rm i}\approx 7.5774045\times 10^{-2}.  
 \end{equation}
 
Now, let us consider the direct cascade. The energy flux per unit of volume is defined as   $P(k)\equiv4\pi  \int_0^k  \kappa^4 \, \text{St}_{\kappa}(x) \, \mathrm{d}\kappa$. 
It appears that this integral is logarithmically divergent for  $x=3/2$, i.e. marginally non-local.
Interestingly, we found $I(3/2)$ is finite but nonzero which means that $n_k \sim k^{-3}$ for a constant direct energy flux obtained by dimensional analysis is not a valid mathematical solution of the WKE and is not physically realizable.
Based on a phenomenological argument analogous to Kraichnan's well-known argument for the log-correction of the direct enstrophy cascade spectrum in the classical 2D turbulence \cite{Kraichnan-inertial}, Refs.~\cite{dyachenko1992optical,nazarenko2011wave} proposed a "log-correction" for $k^{-3}$.
Note that the universal pre-factor constant cannot be determined using such an argument due to its non-rigorous nature.
To address the log-correction systematically,  we introduce an IR cut-off at the forcing wave number $k_{\rm f}$ in the energy flux integral; then $n_k=Ak^{-3}$ leads to
$P(k)=-16\pi^4\,A^3\,I(\tfrac{3}{2})\ln{(\tfrac{k}{k_{\rm f}})}$ for $k>k_{\rm f}$, which is not $k$-independent as  assumed by the KZ spectrum.
Instead, we seek for a solution of the form $n_k=Bk^{-2x}\ln^z{(k^2/k_{\rm f}^2)}$; then
$P(k)$ can be simplified for $k\gg k_{\rm f}$ as
\begin{equation}
P(k)=-16\pi^4B^3\,I(x)\ln^{3z}(\tfrac{k^2}{k_{\rm f}^2})  \medint\int_{k_{\rm f}}^k \kappa^{8-6x} \rmd \kappa\,.
\end{equation}
Constant energy flux requires $x=3/2$ and $z=-1/3$, giving $P_0=-16\pi^4B^3\,I(3/2)$.
Finally, we obtain the log-corrected KZ spectrum for the direct cascade and the universal pre-factor as
 \begin{equation} \label{eq:dc}
 n_{ k} =C_{\rm d} P_0^{1/3}  k^{-3} \ln^{-1/3}\left({k}/{k_{\rm f}}\right)\,,\, C_{\rm d}\approx 5.26\times10^{-2}\,. 
 \end{equation}
  All the details for the derivation of KZ spectra can be found in the SM.

Note that previous GPE numerical simulations in the direct cascade setting \cite{proment2009energy,proment2012sustained},  reported a reasonable agreement with the $-3$ power-law scaling of \eqref{eq:dc}, but the numerical resolution was rather limited and no log-correction was observed or discussed.
In numerical simulations of Ref.~\cite{navon2016emergence}, a steeper scaling with exponent
close to $-3.5$ was reported, which was similar to the experimental result discussed in the same paper. The scale separation there was also relatively modest, 
and no explanation was given for the steeper spectrum. As for the inverse cascade, to date there have been no numerical simulations or experiments done.

We perform numerical simulations of the forced-dissipated GPE using the standard massively-parallel pseudo-spectral code FROST \cite{KrstulovicHDR} with a fourth-order Exponential Runge-Kutta temporal scheme (see \cite{zhu2022testing}). We use grids of $N_p^3$ collocation points, with $N_p=512$ and $N_p=1024$ to verify the numerical convergence.  We add a forcing term $F_{\bf k} (t) $ and a dissipation term $-D_{\bf k}\widehat{\psi}_{\bf{k}} (t) $ to the Fourier transform of the RHS of GPE \eqref{GPE}. The forcing term is supported on a narrow band around the forcing wavenumber $k_{\rm f}$ and it obeys the Ornstein–Ulenbeck process  $\mathrm{d}F_{\bf k} (t)=-\gamma\, \widehat{\psi}_{\bf{k}}\mathrm{d}t+f_0\mathrm{d}{W}_{\bf k}$, where {$W_{\bf k}$} is the Wiener process. The parameters $\gamma$ and $f_0$ control the correlation time and the amplitude of the forcing respectively. Naturally, $k_{\rm f}$ is taken small for the direct cascade and {large for the inverse one}. 
Dissipation is of the form $D_{\bf k} = (k/k_{\rm L})^{-\alpha}+(k/k_{\rm R})^{\beta}$, and acts at small and large scales. Moreover, the condensate ($k=0$ mode) is dissipated in the same manner with a constant friction $D_{\bf 0}$. 
We optimize the parameters of forcing and dissipation in order to enlarge the inertial range for a fixed resolution,
while maintaining simulations well resolved and minimizing bottlenecks at the dissipation scales.
 We pay special attention that forcing is weak enough so that the system fulfills WT assumptions (See SM for verification). Tab.~\ref{tab:param} gives numerical  parameters. 
 Finally, the $k$-space energy and particle fluxes, $P(k)$ and $Q(k)$ respectively, are computed directly using the GPE (\ref{GPE}) (see SM, which includes
Refs. \cite{Griffin2022Energy}).
 \begin{table}[h!]
\begin{ruledtabular}
\begin{tabular}{ccccccc}
case&model&cascade&$L$&$N_p$&$f_0^2$ & $\gamma$\\
\hline
1&\multirow{4}{*}{GPE}&\multirow{2}{*}{direct}&$2\pi$&$512$&$1.2$& \multirow{2}{*}{$20$}\\
2&  &  &$4\pi$&$1024$&$0.1589$ &  \\
3&  &\multirow{2}{*}{inverse}&$2\pi$&$512$&$10^{-4}$ &  \multirow{2}{*}{$0$} \\
4&  &  &$4\pi$&$1024$&$1.26\times10^{-5}$  &   \\
\hline
case & $k_{\rm f}$ & $D_{\bm{0}}$ &$k_L$& $\alpha$ & $k_R$ & $\beta$ \\
\hline
 1, 2&  8&  \multirow{2}{*}{$10^3$}& 2.5 & 2 & 145 & 4  \\
3, 4&  125 &  & 1 & 0.5 & 130 & 6\\
\hline
case & model & cascade & $\omega_{\rm min}$ & $\omega_{\rm max}$ & $\omega_{\rm f}$ & $c_{\rm f}$  \\
\hline
5 &\multirow{2}{*}{WKE}  &direct & $10^{-5}$ & 10 & $3\times 10^{-4}$ & 1 \\
6 & &inverse & 0.1 & $10^5$ & $125^2$ & 50\\
\hline
case & $\Delta\omega_{\rm f}$ &  $\omega_{\rm L}$ & $\alpha$   &  $\omega_{\rm R}$ & $\beta$ & $k_{\rm f}$\\
\hline
5  & $3\times 10^{-4}$ & $10^{-4}$ & 3 & 2 & 4  & $\sqrt{10^{-3}}$\\
6 & 500  & 10 & 4 & $10^5/4.5$ & 7  &$125$
\end{tabular}
\end{ruledtabular}
\caption{\label{tab:param}%
Parameters for GPE and WKE simulations. 
}
\end{table}

We also simulate the WKE with forcing and dissipation  
using the code developed in \cite{SemGreMedNaz,zhu2022testing}. This code solve{s} the WKE expressed in wave-frequency  $\omega$, and uses a decomposition of the integration domain of the RHS of Eq.~\eqref{eq:WKEw} along lines where the integrand has discontinuous derivatives.  
The WKE is solved in the interval $\omega\in[\omega_{\rm min},\omega_{\rm max}]$, and we set $n_{\omega}= n_{\omega_{\rm min}}$ for $\omega<\omega_{\rm min}$, and $n_{\omega}=0$ for $\omega>\omega_{\rm max}$. The WKE is forced by a constant-in-time forcing $f_\omega=c_{\rm f}\, G(\omega)$, where $G(\omega)$ is a Gaussian centered at $\omega_{\rm f}$ and of width $\Delta\omega_{\rm f}$. Dissipation is introduced by adding the term $-[(\omega/\omega_{\rm L})^{-\alpha}+(\omega/\omega_{\rm R})^{\beta}]n_\omega$ to the RHS of WKE. 
 For time integration, we use a new approach inspired by Chebyshev interpolation and schemes described in \cite{Sem}.
 Values of the parameters are reported in Tab.~\ref{tab:param}. In this Letter, we present solutions of the WKE in $k$-variables to simplify comparisons with GPE data. Standard WKE-based $k$-dependent fluxes are given in the SM.

 \begin{figure}[h!]
  \centering
  \includegraphics[scale=1.]{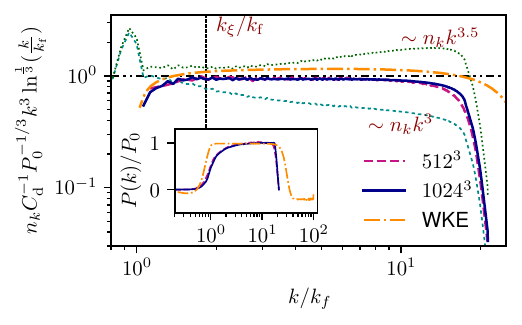}
  \caption{\label{fig:direct}Wave action spectra for the direct cascade compensated by theoretical prediction (\ref{eq:dc}). Data obtained by GPE at two different resolutions and by WKE, respectively.
  Insets: corresponding energy fluxes normalized by their values measured in inertial range.}
  \end{figure}
 First, we present numerical results for the direct cascade state.  Figure \ref{fig:direct} 
 displays the stationary wave action spectra obtained in simulations of the GPE and the WKE respectively, both compensated by the theoretical prediction (\ref{eq:dc}). 
The insets show their respective scale-dependent energy fluxes, normalized by $P_0$ measured in the range where  $P(k)$ is approximately constant. 
The same values of $P_0$ are used in \eqref{eq:dc}.
 The values of $k_{\rm f}$, as presented in Tab.~\ref{tab:param}, are selected within the range of forcing.
For comparison, we also plot the compensated KZ spectra $\sim n_k k^3$ (ignoring the non-locality issue) and $\sim n_k k^{3.5}$ for the GPE data with $N_p=1024$.
We see an excellent agreement between \eqref{eq:dc} and GPE and WKE data including the value of the constant $C_{\rm d}$.
The vertical dotted line denotes the wave vector $k_\xi$ where the nonlinear term in the GPE becomes
equal  to the linear one. WT prediction is expected to be valid at $k>k_\xi$ only. 
Further,  the asymptotic result \eqref{eq:dc} is  assumed for $k\gg k_{\rm f}$. Interestingly, the theoretical log-corrected KZ spectrum provides a very good fit to the numerical results even at the scales $k \lesssim k_{\xi} \sim k_{\rm f}$.
Note that GPE data with $N_p=1024$  present a relatively good agreement with $k^{-3.5}$ too, although in a much narrower range and only at low $k$, which is consistent with the results reported in \cite{navon2016emergence}.

\begin{figure}[h!]
  \includegraphics[scale=1.]{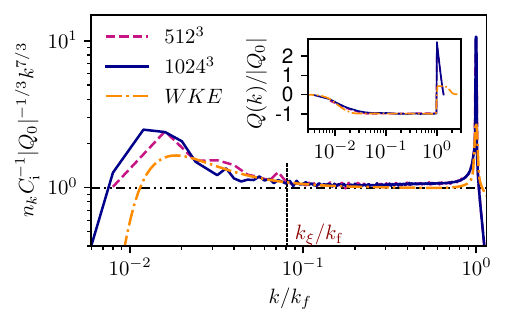}
  \caption{\label{fig:inverse}Wave action spectra for the inverse cascade compensated by theoretical prediction (\ref{eq:ic}). Data obtained by GPE at two different resolutions and by WKE, respectively.
  Insets: corresponding particle fluxes normalized by their values measured in inertial range. }
  \end{figure}

 Next, we study the inverse cascade state. Figure \ref{fig:inverse}
 shows the wave action spectra and the particle fluxes $Q(k)$ (on insets normalized by $|Q_0|$) obtained in GPE and WKE simulations.
 Spectra are compensated by the theoretical prediction \eqref{eq:ic} including the value of the pre-factor $C_{\rm i}$.
 Again, for GPE data we mark $k_\xi$ by a vertical dotted line. For both GPE and WKE we see a significant range (within the constant-$Q$ region) where the compensated spectra have plateaus, which confirms the predicted spectrum \eqref{eq:ic}. The agreement between theory and numerics is almost perfect for WKE data and within $5\%$ for GPE. Note that in both simulations we see a "bump" on the left part of the spectrum, which could be attributed to an infrared bottleneck caused by the nature of the hypo-viscous dissipation.

Finally, to check the reliability of our predictions in a setting closer to experiments, we study the direct and inverse cascades for BEC trapped in a cubic box. To this end, while solving \eqref{GPE} we consider a trapping potential $U({\bf x})$ that vanishes inside the box and increases rapidly at the borders of the cube (see SM for an exact definition).
\begin{figure}[h!]
  \includegraphics[width=.99\linewidth]{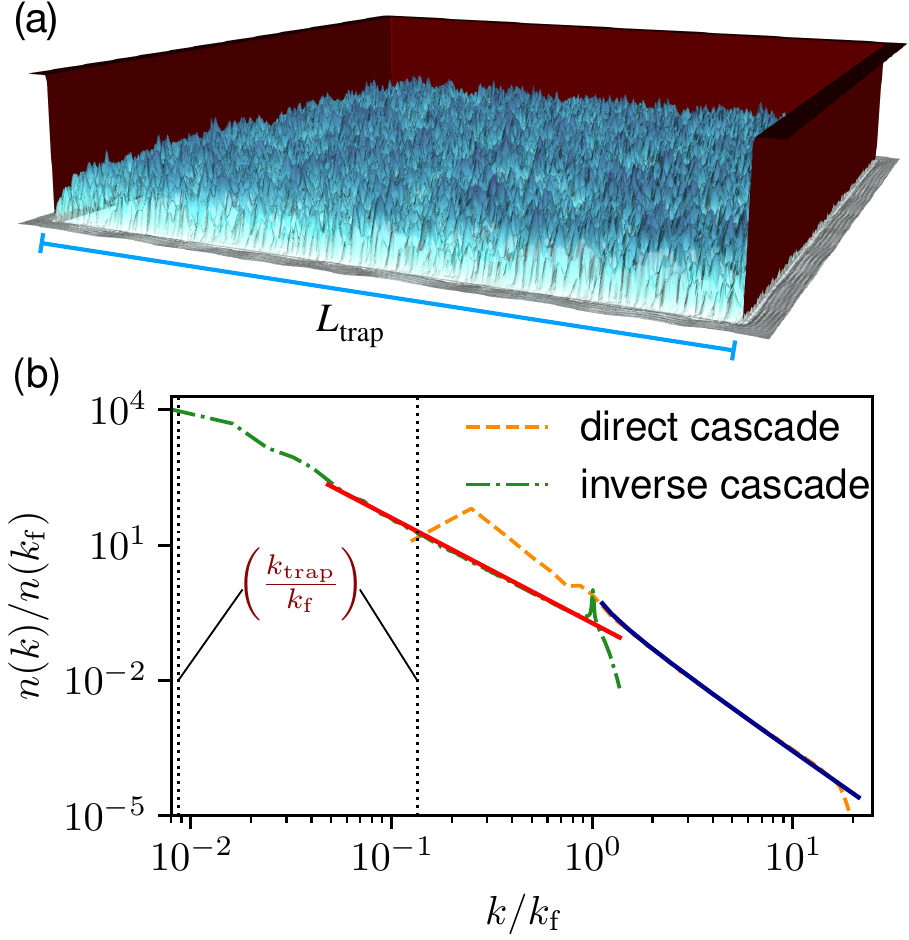}
  \caption{\label{fig:pot} GPE simulations of turbulent trapped BEC. (a) Two-dimensional cut of a typical simulation (arbitrary units and scales). The trapping potential is displayed in red and density fluctuations in blue. (b) Wave action spectra normalized by {their values} at the respective forcing scales $k_{\rm f}$ for the direct and the inverse cascades. The solid lines display the theoretical predictions (\ref{eq:ic},\ref{eq:dc}).
  The vertical lines show the wave number $k_{\rm trap}=2\pi/L_{\rm trap}$for both of the cascades}, with $L_{\rm trap}$ the trap size. 
  \end{figure}
Figure \ref{fig:pot} (a) displays a two-dimensional cut of a typical simulation where we plot the trapping potential and the wave-field. We keep the same forcing and dissipation schemes and the parameters of case 1 and case 3 in Tab.~\ref{tab:param} respectively. The results for both the direct and the inverse cascades are shown in Fig.~\ref{fig:pot}, superimposed with the theoretical KZ spectra (solid lines) without any fitting parameters. Once again, one can see a nearly perfect agreement, which indicates robustness of our theoretical predictions and their relevance to the past and future experiments on BEC turbulence. It might be convenient for comparison with experiments to rewrite our predictions (\ref{eq:dc}, \ref{eq:ic}) in  dimensional form. In terms of the reduced Planck constant $\hbar$, the interaction constant $g$ and boson mass $m$, they read (see SM)
\begin{eqnarray}
 {\rm direct}: &\quad&n_{ k} =C_{\rm d} \left(P_0\hbar/g^2\right)^{1/3}  k^{-3} \ln^{-\frac{1}{3}}\left({k}/{k_{\rm f}}\right)\! ,\label{Eq:DirectDimensional}\\
{\rm inverse}: &\quad& n_{ k} =C_{\rm i} \left(|Q_0|\hbar^3/2g^2m\right)^{1/3} k^{-7/3}\,.\label{Eq:InverseDimensional}
\end{eqnarray}
Note the $n_k$ is dimensionless and normalized such that $N=\int n_k \mathrm{d}^3{\bf k}$ is the total number of particles per unit of volume.

 Summarizing, in this Letter we have derived the stationary direct and inverse cascade KZ spectra (\ref{eq:ic}, \ref{eq:dc}), including, for the first time, the analytical determination of the logarithmic correction  in \eqref{eq:dc} and the pre-factor constants for both. Our predictions are in remarkable agreement, without any adjustable parameters, with numerical simulations of the GPE and the WKE.  Such definitive agreement was possible thanks to considerable higher than in the previous works resolution of the GPE simulations and careful checks of the WT assumptions. 
 To our knowledge, we also presented the first simulations of the associated WKE in the steady state regimes. 
 
 In the case of the direct cascade, previous works reported a steeper $-3.5$ exponent \cite{navon2016emergence} -- a result which deviates from the dimensional WT prediction $-3$.
Several processes (a residual role of vortices, the non-negligible incompressible-flow energy, 
an increasing importance of quantum pressure)  were suggested in \cite{navon2016emergence} as candidates for explaining the experimental result but without  an argument why they could lead to the $k^{-3.5}$ spectrum.
We do not think that such additional processes are important and/or need to be considered because the logarithmic correction  in Eq. \eqref{eq:dc} was derived without invoking other physical phenomena than weak wave turbulence. It is convincingly confirmed by our high-resolution numerical results and it agrees with the experimental spectrum reported in \cite{navon2016emergence} (see SM).
 Hence, we conclude that the spectrum $k^{-3.5}$ is an
 approximation to the log-corrected KZ spectrum.
 Further, for the first time, in our work the inverse cascade KZ spectrum is observed numerically.
 
  Our results are useful for laboratory experiments, and we have validated them with simulations of GPE of a trapped BEC. In future, it would be particularly interesting to have a stationary inverse cascade state experimentally implemented. For this, one could use a similar forcing technique as in \cite{navon2016emergence,navon2019synthetic}, namely shaking the retaining trap. In addition, one would have to devise a synthetic dissipation mechanism removing low momentum atoms which would prevent their accumulation (condensation) near the ground state of the trap thereby making a statistically steady state possible.

\begin{acknowledgments}
This work was funded by the Simons Foundation Collaboration grant Wave Turbulence (Award ID 651471). 
This work was granted access to the high-performance computing facilities under GENCI (Grand Equipement National
de Calcul Intensif) A0102A12494 (IDRIS and CINES), the
OPAL infrastructure from Université Côte d’Azur, supported
by the French government, through the UCAJEDI Investments in the Future project managed by the National Research
Agency (ANR) under Reference No. ANR-15-IDEX-01, and
the SIGAMM infrastructure (http://crimson.oca.eu)  hosted by Observatoire de
la Côte d’Azur, supported by the Provence-Alpes Côte
d’Azur region and supported by the state contract of the Sobolev Institute of Mathematics (project no. FWNF-2022-0008)
\end{acknowledgments}

\bibliographystyle{apsrev4-1}
\bibliography{WWT}

\clearpage

\appendix
\begin{widetext}
  \begin{center}
    \Large
    \textbf{Direct and inverse cascades in turbulent Bose-Einstein condensate: Supplemental material}
  \end{center}

  \renewcommand\thefigure{S\arabic{figure}}
  \setcounter{figure}{0}
  \setcounter{subsection}{0}
  \setcounter{secnumdepth}{2}
  \renewcommand{\thesubsection}{\Roman{subsection}}
  \titleformat{\subsection}{\large\bfseries\centering}{\thesubsection .}{0.5em}{}
  \setcounter{equation}{0}
  \renewcommand\theequation{A\arabic{equation}}
    
\subsection{Behavior of the collision term on the power-law spectra \label{sec:Ci}}

We consider the wave kinetic equation \eqref{WKE} expressed in wave-frequency variables. It reads
\begin{equation}\label{eq:WKEw}
  \frac{\partial n_{\omega}}{\partial t} =St(\omega,t) \equiv \frac{4\pi^3}{\sqrt{\omega}}\int \left[\min\left(\omega,\omega_1,\omega_2,\omega_3\right)\right]^{1/2}
  n_{\omega}  n_{\omega_1}n_{\omega_2}n_{\omega_3} \left(\frac{1}{n_{\omega}} + \frac{1}{n_{\omega_1}} - \frac{1}{n_{\omega_2}}-\frac{1}{n_{\omega_3}}\right)\delta(\omega^{01}_{23}) \rmd\omega_1\rmd\omega_2\rmd\omega_3,  
  \end{equation}
  where now $\omega^{01}_{23}=\omega+\omega_1-\omega_2-\omega_3$.

The mean density of particles and 
the mean density of energy in terms of $\omega$ become
\begin{subequations}
\begin{equation}\label{eq:conserv_N_H}
   N=2\pi\int_0^\infty \omega^{1/2} n(\omega,t)\rmd\omega\,,  \qquad 
  H=2\pi\int_0^\infty \omega^{3/2} n(\omega,t)\rmd\omega \,.
   \tag{\theequation a-b}  
\end{equation}
\end{subequations}
Consequently, the particle and energy scale-dependent fluxes are defined in the standard way \cite{nazarenko2011wave},
\begin{subequations}
\begin{equation}\label{eq:PQ}
 Q(\omega,t)=-2\pi\int_0^\omega \tilde{\omega}^{1/2} St(\tilde{\omega},t)\rmd\tilde{\omega}  \,, \qquad
 P(\omega,t)= -2\pi\int_0^\omega \tilde{\omega}^{3/2} St(\tilde{\omega},t)\rmd\tilde{\omega}  \,. 
   \tag{\theequation a-b}  
\end{equation}
\end{subequations}

Let us substitute a power-law spectrum $n_{\omega}=A\omega^{-x}$ (not necessarily a WKE solution) into the WKE \eqref{eq:WKEw} and rewrite the later as follows,
\begin{equation}\label{eq:WKE}
\frac{\partial n_{\omega}}{\partial t} = 4\pi^3\,A^3\,\omega^{-3x+2}\,I(x)\,,
\end{equation}
with the dimensionless collision term 
 \begin{equation} 
 \label{eq:Ix}
I(x)=\int \left[\min\left(1,q_1,q_2,q_3 \right)\right]^{1/2} 
\left( q_1q_2q_3 \right)^{-x}
\left(1+q_1^x-q_2^x-q_3^x\right)
 \delta \left(q^{01}_{23}\right)
 \rmd q_1 \rmd q_2 \rmd q_3\,,
\end{equation} 
where we performed the change of variables $q_i=\omega_i/\omega$ for $i=1,2,3$, and now $\delta \left(q^{01}_{23}\right)=\delta\left(1+q_1-q_2-q_3\right)$. The integral is taken over $q_1\,,q_2\,,q_3>0$. 
If we integrate over $q_1$, the integration domain in $(q_2, q_3)$-plane becomes the following: $q_2,q_3>0\,,q_2+q_3-1=q_1>0 $. 
One can apply the Zakharov transformation (ZT) \eqref{ZT} to $I(x)$ and get
\begin{equation} \label{eq:Iztx}
I_{\rm ZT}(x)=\int q_1^{1/2-x}
\left(q_2q_3 \right)^{-x}
\left(1+q_1^x-q_2^x-q_3^x\right)
\left(1+q_1^y-q_2^y-q_3^y\right)
 \delta \left(q^{01}_{23}\right)
 \rmd q_1 \rmd q_2 \rmd q_3\,,
\end{equation}
with $y=3x-7/2$, and the integral domain is changed to $0<q_1\,,q_2\,,q_3<1$. Note that $I(x)=I_{\rm ZT}(x)$ only if the integrals are convergent.

 \begin{figure}[h]
\centering
\includegraphics{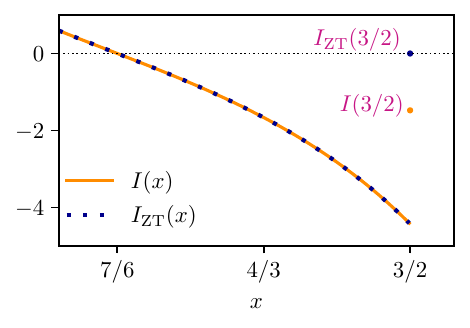}%
\caption{\label{figI} $I(x)$ and $I_{\rm ZT}(x)$ in the window of convergence. 
}
\end{figure}
Function $I_{\rm ZT}(x)$ has two zeros:  $x=3/2$ ($y=1$) corresponding to the forward cascade of energy $n_{\omega}=A\,\omega^{-3/2}$  and  $x=7/6$ ($y=0$) --- to the inverse cascade of particles $n_{\omega}=A\,\omega^{-7/6}$. However, the ZT is not an identity transformation and, therefore, 
these candidates to the stationary solutions must be checked by substituting them into the original integral 
$I(x)$ and making sure that the resulting integral is convergent and equal to zero. Physically, such an integral convergence means that  wave quartets with similar values of the frequencies dominate the nonlinear evolution; hence this property is called the interaction locality. Mathematically, violation of locality (convergence) simply means that the considered spectrum is not a valid solution.

It was shown in \cite{SemGreMedNaz} that $I(x)$ is convergent for
$1<x<3/2$. 
Fig.~\ref{figI}  plots  $I(x)$ and $I_{\rm ZT}(x)$ calculated numerically for $x\le3/2$. It is easy to see that the two integrals do coincide in the interval $1<x<3/2$. Therefore, 
the inverse cascade spectrum $n_\omega=A\omega^{-7/6}$ is local, and it is a valid mathematical solution of the WKE.
It is also interesting to note that $I(3/2)$ is actually convergent, and its value can be  derived analytically (see Sec. \ref{sec:Cd}).
The finite non-zero value of $I(3/2)$ implies that, although the collision integral of the WKE is convergent for $x=3/2$, 
the power-law (direct cascade) spectrum with this exponent is not an exact stationary solution of the WKE. 
However, with a logarithmic correction this spectrum can be made a valid asymptotical solution for the direct cascade (see Sec. \ref{sec:Cd}). 

\subsection{Derivation of the Kolmogorov-Zakharov constant $C_{\rm i}$ for the inverse cascade}

In order to get the KZ constant for the inverse cascade, we rewrite the WKE in the particle conservation form:
\begin{equation}\label{eq:WKE1}
\frac{\partial\left( 2\pi\omega^{1/2}n_{\omega}\right)}{\partial t} = -\frac{\partial Q(\omega, t)}{\partial \omega}\equiv 8\pi^4\,A^3\,\omega^{-3x+5/2}\,I_{\rm ZT}(x)\,.
\end{equation}
Here we replace $I(x)$ with $I_{\rm ZT}(x)$ for simplicity since that the two are equal for $1<x<3/2$.
The particle flux by definition is 
\begin{equation}\label{eq:Qomega}
    Q(\omega,t)=\int_0^{\omega}\left( 
    -8\pi^4\,A^3\,\tilde{\omega}^{-3x+5/2}\,I_{\rm ZT}(x)
    \right)d\tilde{\omega} 
    =8\pi^4 A^3\omega^{-y}\frac{I_{\rm ZT}(x)}{3x-7/2}\,.
\end{equation}
In the limit $y\to0\,,x\to 7/6$, when the wave system goes to the stationary state with a constant (frequency-independent) $Q$,   by the L'Hopital rule, we obtain   $Q=8\pi^4A^3I'_{ZT}(7/6)/3$, 
where prime stands for the $x$-derivative.
This leads to 
\begin{equation}\label{eq:kzi}
n_{\omega}=3^{1/3}\left(8\pi^4I'_{ZT}(7/6)\right)^{-1/3}Q^{1/3}\omega^{-7/6}.
\end{equation}
Calculating the $x$-derivative, we find
\begin{equation} 
I'_{ZT}(7/6) =3\, \int \limits_{0<q_1,q_2,q_3<1}  q_1^{-2/3}
\left(q_2q_3 \right)^{-7/6}
\left(1+q_1^{7/6}-q_2^{7/6}-q_3^{7/6}\right)
\ln\frac{q_1}{q_2q_3}
 \delta \left(q^{01}_{23}\right) 
 \rmd q_1 \rmd q_2 \rmd q_3\,,
\end{equation}
which value can be computed using Mathematica. 
Passing \eqref{eq:kzi} to wave number variable $k$, we get $n_k=C_{\rm i} |Q|^{1/3} k^{-7/3}$, where  $C_{\rm i}$ reads
 \begin{equation}
 \begin{split} \label{eq:ci}
 C_{\rm i}=&\frac{1}{2\pi^{3/2}}\Gamma\left(\tfrac{5}{6}\right)^{1/3}
 \Bigg[ 3\Gamma \left(\tfrac{1}{3}\right) 
 \Big(  3^{3/2}\,2^{2/3}\,_3F_2\left(\tfrac{1}{6},\tfrac{1}{6},\tfrac{1}{3};\tfrac{4}{3},\tfrac{4}{3};1\right) 
  -8 \,_3F_2\left(\tfrac{1}{6},\tfrac{1}{3},\tfrac{1}{3};\tfrac{4}{3},\tfrac{3}{2};1\right)\\
 &  +2^{1/3} \, _3F_2\left(\tfrac{1}{3},\tfrac{1}{3},\tfrac{1}{2};\tfrac{3}{2},\tfrac{5}{3};1\right) 
   -2^{1/3} \,_4F_3\left(\tfrac{1}{3},\tfrac{1}{3},\tfrac{1}{2},\tfrac{1}{2};\tfrac{3}{2},\tfrac{3}{2},\tfrac{5}{3};1\right
   )   \Big)\Bigg]^{-1/3}\, \approx 7.5774045\times 10^{-2},
\end{split}
\end{equation}
where $\Gamma(\cdot)$ is the Gamma function, and $_pF_q\left({a_1,\dots,  a_p};{b_1,\dots,b_q};z\right)$ is the generalized hyper-geometric function.

\subsection{Derivation of the direct energy cascade Kolmogorov-Zakharov spectrum}  \label{sec:Cd}

To illustrate the mathematical issues of the direct energy cascade KZ solution, we first assume a pure power-law form of the spectrum. We rewrite the WKE in the energy conservation form:
\begin{equation}\label{eq:WKE1}
\frac{\partial\left( 2\pi\omega^{3/2}n_{\omega}\right)}{\partial t} = -\frac{\partial P(\omega, t)}{\partial \omega}\equiv 8\pi^4\,A^3\,\omega^{-3x+7/2}\,I(x)\,,
\end{equation}
where the original collision integral  $I(x)$ is kept instead of $I_{\rm ZT}(x)$ because of the departure of these two at $x=3/2$, the value corresponding to the direct energy cascade. Then, the energy flux is by definition 
\begin{equation}\label{eq:Pomega}
    P(\omega,t)=\int_0^{\omega}\left( 
    -8\pi^4\,A^3\,\tilde{\omega}^{-3x+7/2}\,I(x)
    \right)\rmd \tilde{\omega} \,.
\end{equation}

As illustrated in Fig \ref{figI},
$I(3/2) \neq 0$.
Actually, it can be computed using Mathematica, which gives 
\begin{equation} \label{eq:I3_2}
I(3/2)= \int\limits_{q_1,q_2,q_3>0}
\min\left(1,q_1,q_2,q_3 \right)^{1/2} 
\left( q_1q_2q_3 \right)^{-3/2}
\left(1+q_1^{3/2}-q_2^{3/2}-q_3^{3/2}\right)
 \delta \left(q^{01}_{23}\right)
 \rmd q_1 \rmd q_2 \rmd q_3
    = -4 \pi + 16 \ln{2}\,.
\end{equation}
The fact that $I(x)$ is discontinuous at $3/2$, and that the value of either $I(3/2)$, or  $I(3/2^-)$ --- equal to $I_{\rm ZT}(3/2^-)$, where $x=3/2^-$ means the limit  $x\to 3/2$ taken from below (see Fig. \ref{figI}), -- is nonzero and finite results in inapplicability of  the L'Hopital rule for calculating the limit $x\to 3/2$ (supposing $x<3/2$),
i.e. we cannot use the same procedure to compute $C_{\rm d}$ as we used before for the calculation of $C_{\rm i}$. Moreover, Eq.~(\ref{eq:Pomega}) gives a logarithmically divergent energy flux integral after substituting $x=3/2$. 
The divergence of flux at finite frequency can be avoided by cutting the 
integral off at the forcing frequency $\omega_{\rm f}=k_{\rm f}^2$, which leads to
\begin{equation}\label{eq:Pomega1}
    P=\int_{\omega_{\rm f}}^{\omega}\left( 
    -8\pi^4\,A^3\,\tilde{\omega}^{-1}\,I(3/2)
    \right)\rmd\tilde{\omega} 
    =-8\pi^4\,A^3\,I(3/2)\ln{\frac{\omega}{\omega_{\rm f}}}\,.
\end{equation}
However, the  flux $P$ must be independent of $\omega$ for steady state solutions.
This is clearly not the case in the above expression, which is another
indication that $n_{\omega}\sim\omega^{-3/2}$ is not a valid stationary solution of the WKE.

%
To remove the $\omega$-dependence term $\ln{\frac{\omega}{\omega_{\rm f}}}$ in (\ref{eq:Pomega1}), we introduce a logarithmic correction and seek solution as 
$n_\omega=C\omega^{-x}\ln^z{\frac{\omega}{\omega_{\rm f}}}$.
Applying the cut-off at $\omega_{\rm f}$, the energy flux becomes
\begin{equation}\label{eq:[p,ega2}
\begin{split}
P=&-8\pi^4C^3\int_{\omega_{\rm f}}^\omega \tilde{\omega}^{-3x+7/2}d\tilde{\omega} \int\limits_{q_1,q_2,q_3>\tfrac{\omega_{\rm f}}{\tilde{\omega}}}
\min\left(1,q_1,q_2,q_3 \right)^{1/2} 
\left( q_1q_2q_3 \right)^{-x}  \delta \left(q^{01}_{23}\right) \\
&\left(\ln^{-z}\tfrac{\tilde{\omega}}{\omega_{\rm f}}+q_1^x\ln^{-z}\tfrac{\omega_1}{\omega_{\rm f}}-q_2^x\ln^{-z}\tfrac{\omega_2}{\omega_{\rm f}}-q_3^x\ln^{-z}\tfrac{\omega_3}{\omega_{\rm f}}\right)\ln^z\frac{\tilde{\omega}}{\omega_{\rm f}}\ln^z\frac{\omega_1}{\omega_{\rm f}}\ln^z\frac{\omega_2}{\omega_{\rm f}}\ln^z\frac{\omega_3}{\omega_{\rm f}}
 \rmd q_1 \rmd q_2 \rmd q_3\,.
\end{split}
\end{equation}
For $\omega \gg \omega_{\rm f}$, we note that the main contribution comes from $\tilde{\omega}\,,\omega_1\,,\omega_2\,,\omega_3\gg\omega_{\rm f}$; then $\ln\frac{\tilde{\omega}}{\omega_{\rm f}}\approx\ln\frac{\omega_1}{\omega_{\rm f}}\approx \ln\frac{\omega_2}{\omega_{\rm f}}\approx  \ln\frac{\omega_3}{\omega_{\rm f}}\approx \ln\frac{\omega}{\omega_{\rm f}}$. Thus, we obtain
\begin{equation}
\begin{split}
P=&-8\pi^4C^3\,I(x)\ln^{3z}\frac{\omega}{\omega_{\rm f}}  \int_{\omega_{\rm f}}^\omega \tilde{\omega}^{-3x+7/2} \rmd \tilde{\omega}\,.
\end{split}
\end{equation}
Note that we have replaced the lower limits in the second integral in \eqref{eq:[p,ega2} by zero because the resulting integral is convergent.
The independence  of $P$ from $\omega$ requires $x=3/2$ and $z=-1/3$, which leads to 
$n_\omega =\left(-8\pi^4I(3/2)\right)^{-1/3}P^{1/3}\omega^{-3/2} \ln^{-1/3}(\omega/\omega_{\rm f}) \,.$
Then, in terms of $k$ we have 
\begin{equation} \label{eq:KZd}
n_k =\left(-16\pi^4I(3/2)\right)^{-1/3}P^{1/3}k^{-3} \ln^{-1/3}(k/k_{\rm f}) \,.
\end{equation}
It gives us the log-corrected KZ spectrum 
$n_k =C_{\rm d} P^{1/3}  k^{-3} \ln^{-\frac{1}{3}}\left({k}/{k_{\rm f}}\right)$ with
\begin{equation}\label{eq:Cd}
C_{\rm d} =\left(-16\pi^4I(3/2)\right)^{-1/3} \,.
\end{equation}
The  analytical expression $I(3/2)=-4 \pi + 16 \ln{2}$ gives  $C_{\rm d}\approx7.58\times 10^{-2}$, 
and numerically obtained $I(3/2^-)\approx-4.42$ gives another possible approximate value for $C_{\rm d}$ as $5.26\times 10^{-2}$. In fact, we will use this latter value as it appears to agree perfectly with our numerical simulations. 

Once again, it should be emphasized that the spectrum by (\ref{eq:KZd}) is supposed to be valid only for $\omega\gg\omega_{\rm f}$.

\subsection{Energy and particle fluxes for the Gross-Pitaevskii equation}
According to the definition of the mean energy density \eqref{eq:cons-H}, we
write the  energy flux $P(k)$  as
\begin{equation}
P(k)=-\Delta_k\sum_{p=0}^k \left.\frac{\partial H(p)}{\partial t}\right|_{\rm Ham}=-\Delta_k\left(\sum_{p=0}^k \left.\frac{\partial H_2(p)}{\partial t}\right|_{\rm Ham} +\left.\frac{\partial H_4(p)}{\partial t}\right|_{\rm Ham}  
\right)\,,
\end{equation}
where $H_2(k)$ is the energy spectrum of the quadratic part (with respect to $\psi$) that coincides with the 1D energy spectrum $E(k)$ defined in the 
WT theory,  $H_4(k)$ is the 1D spectrum of the fourth-order  part of the energy that is not computed in the WT theory,
$\Delta_k=2\pi/L$ is the mesh size in Fourier space. The vertical bars $\left.\right|_{\rm Ham}$ mean that time derivatives are taken only of the Hamiltonian part of the GP equation (without forcing and dissipation).
Consequently, we have (see \cite{Griffin2022Energy})
\begin{equation}
  \left.\frac{\partial H_2(k)}{\partial t}\right|_{\rm Ham}=2 E_{\nabla \psi,\nabla \dot{\psi}}(k)\,,\quad
  \left.\frac{\partial H_4(k)}{\partial t}\right|_{\rm Ham}= E_{\rho,\dot{\rho}}(k)\,,
\end{equation}
where $\rho=\psi\,\psi^*$, $\dot{\rho}=\psi \dot{\psi}^*+\dot{\psi} \psi^*$, with  
$\dot{\psi}=i\left[ \nabla^{2}  -|\psi|^{2} \right]\psi$ the Hamiltonian part in the RHS of the GPE. The cross spectrum of two fields $f$ and $g$ is defined in terms of their Fourier transform $\hat{f}$ and $\hat{g}$ as 
$$E_{f,g}(k)=\frac 1{\Delta_k} \sum_{k-\Delta_k/2<|\bm{k}|<k+\Delta_k/2} \hat{f}_{\bm{k}}\hat{g}_{\bm{k}}^*\,.$$
Similarly, one can find the particle flux as 
\begin{equation}
Q(k)=-\Delta_k\sum_{p=0}^k \frac{\partial n(p)}{\partial t}=-2\Delta_k E_{\psi,\dot{\psi}}(k)\,.
\end{equation}
It is worth noting that  $\lim_{k\to\infty}P(k)=0$ and $\lim_{k\to\infty}Q(k)=0$ because of the energy and particle conservation of the GPE. Obviously, in numerics the infinity must be replaced with the maximum wave number, $k_{max}$.

\subsection{Verifying the assumptions of Wave Turbulence theory for the Gross-Pitaevskii simulations} 

\begin{figure*}[h]
\centering
\includegraphics{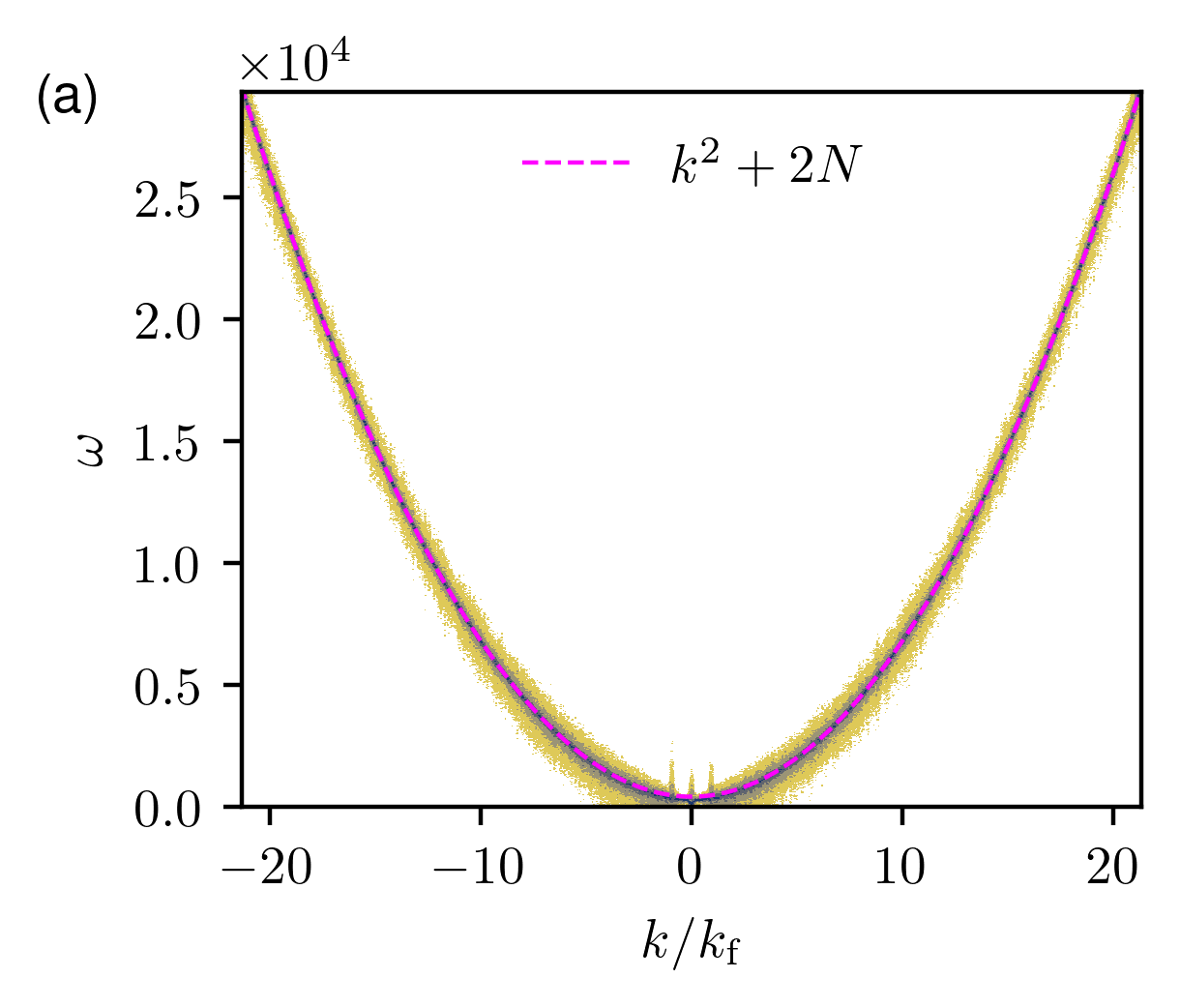}%
\hfill
\includegraphics{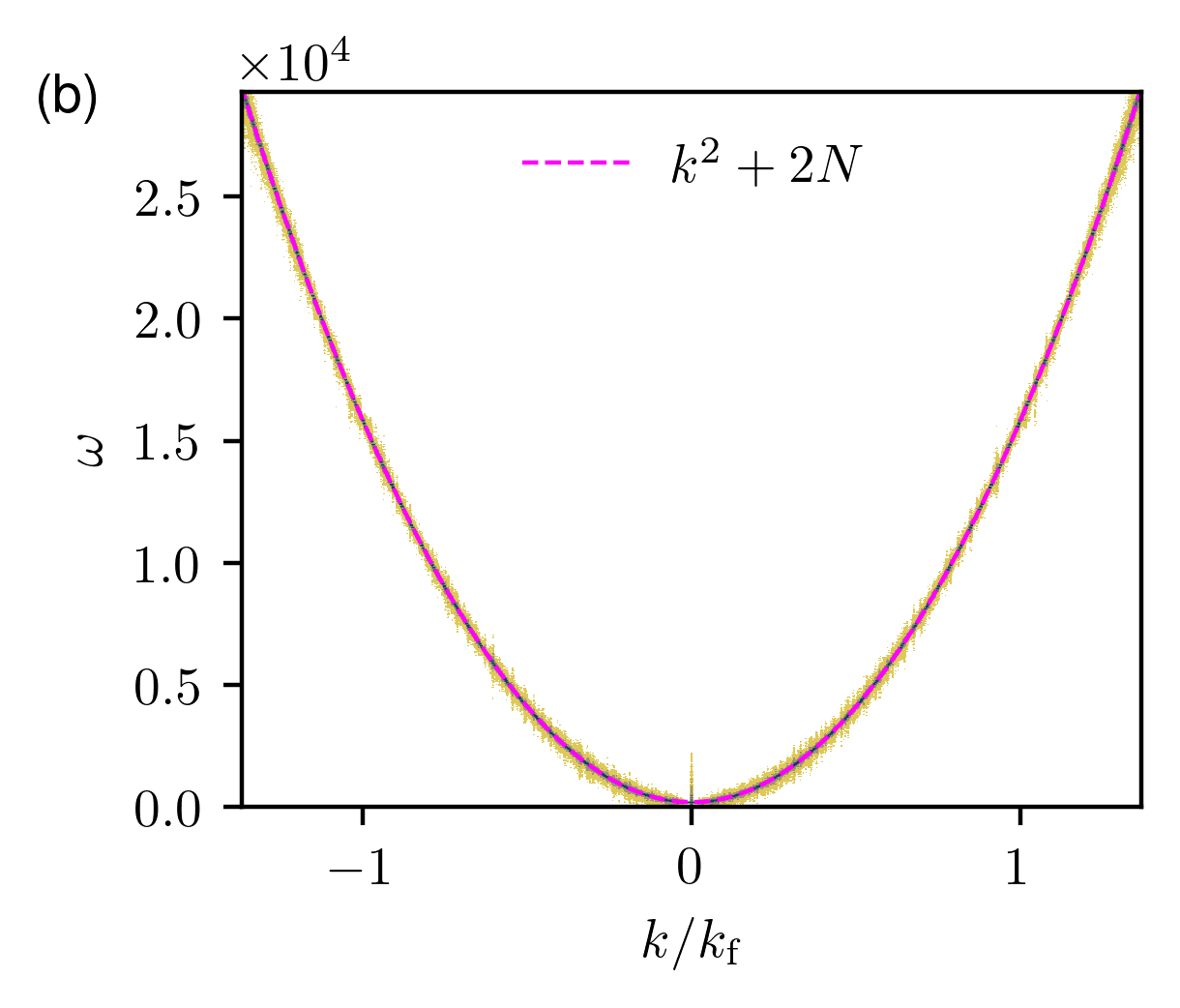}
\caption{\label{fig:ST}
Normalised spatio-temporal spectral density of $\psi(\bm{r}, t)$ for (a) the direct cascade; (b) the inverse cascade. 
}
\end{figure*}
Fig.~\ref{fig:ST} displays the normalized spatio-temporal spectral density $S(\omega,k)\propto |\hat{\psi}(k,\omega)|^2$, where $\hat{\psi}(k,\omega)$ is the time and space Fourier transform of $\psi(\bm{r}, t)$, for the GPE simulations with resolution of $1024^3$ of the direct and the inverse cascades respectively. The time Fourier transform was performed over a time window $T$, which is large enough so that the window-related spectrum broadening is significantly less than the broadening due to the nonlinear effects. We used $T=0.75$ and $1$ for the direct and inverse case respectively.
The figure demonstrates that most of the spectrum is concentrated near the frequencies that satisfy the dispersion relation 
$\tilde{\omega}(k)=k^2 + 2N$  with  $\omega(k)=k^2$ the
linear-wave dispersion relation, and $2N$  the shift induced by the nonlinearity.
 
\begin{figure*}[h]
\centering
\includegraphics{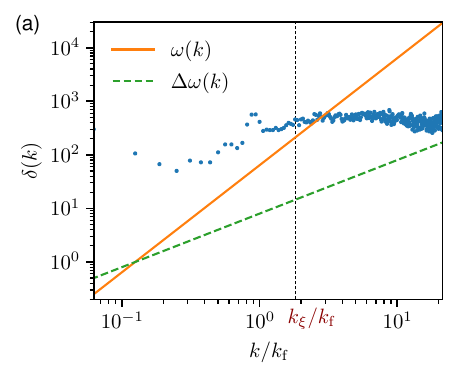}%
\hfill
\includegraphics{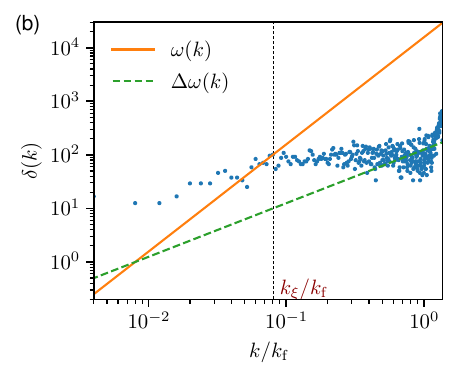}
\caption{\label{fig:deltak} 
Frequency broadening $\delta(k)$ (blue points)  obtained from the spatial-temporal spectral density for (a) the direct cascade; (b) the inverse cascade.
}
\end{figure*}

Further, the frequency broadening $\delta(k)$ measured around $\tilde{\omega}(k)$ from the spatial-temporal spectral density is presented 
in Fig. \ref{fig:deltak}. For applicability of the WT theory, the frequency broadening, caused by the nonlinearity, should be  sufficiently narrow, $\delta(k)<\omega(k)$, so that the nonlinear time 
scales are much greater than the linear ones. This condition ensures that the waves are weakly nonlinear. 
On the other hand, for the solutions of
the GPE in the discrete Fourier space to be in the continuous k-space regime assumed by the WT theory, $\delta(k)$ should be greater than the 
frequency distance between the adjacent wave modes $\Delta\omega(k)=2k\Delta k$. 
The $k$-ranges that satisfy the  two constraints of WT theory, as shown in Fig. \ref{fig:deltak}, are $2k_{\rm f}$--$20k_{\rm f}$ for the direct cascade 
and $0.06k_{\rm f}$--$k_{\rm f}$ for the inverse cascade respectively. These ranges are consistent with those where expected KZ spectra are observed in the main text.  
Details on how the normalised spatio-temporal spectral density and frequency broadening are calculated can be found in \cite{zhu2022testing}. 

\subsection{Potential trap profile}

Ideally, we would like add potential trap such that $U({\bf x})=0$ for ${\bf x}=(x_1\,,x_2\,,x_3)$ inside the trap and $U_0$ elsewhere. Instead, in order to properly use a pseudo-spectral code and avoid spurious Gibbs oscillation, we add a smoothed potential using hyperbolic tangents and periodic functions. Its explicit form is  


\begin{equation}\label{eq:Ubis}
  U({\bf x})=U_0\min{\left\{1,\frac{3}{2}+\frac{1}{2}\sum\limits_{i=1}^3 \tanh{\left[\frac{\tilde{x}_i^2-\left(L_{\rm trap}/3\right)^2}{2\Delta s^2}\right]}  \,\right\}}
\end{equation}
where $\tilde{x}_i=\tfrac{L}{\pi} \sin\tfrac{(x_i-L/2)\pi}{L}$ is a periodic approximation of the the function $x_i-L/2$ near $L/2$, where $L$ is the size of the computational periodic box. The parameter $\Delta s$ smoothens the transition from $0$ to $U_0$ and $L_{\rm trap}$ is the resulting trap size. In numerical simulations we use $U_0=5\times10^4$, $L_{\rm trap}=0.95L$ and $\Delta s=2 \Delta x$, with $ \Delta x =L/N_p$ the mesh size.

\subsection{Nondimensionalization of the Gross-Pitaevskii equation}

The Gross-Pitaevskii equation is often written in a dimensional form and expressed in terms of physical constants as
 \begin{equation}
   i\hbar\frac{\partial\psi ({\bf x},t)  }{\partial t}= -\frac{\hbar^2}{2m}\nabla^{2}\psi  +g\left|\psi \right|^{2}\psi \,,
  \label{GPEd}
 \end{equation} 
where $\hbar$ is the reduced Planck constant, $m$ is the mass of the fundamental bosons and $g=4\pi \hbar^2 a_s/m$ is the coupling constant, with $a_s$  the  boson–boson $s$-wave scattering length. 
The re-scaling $t=\tau t'$, $x=\lambda x'$ and $\psi=\lambda^{-3/2}\psi'$, with $\tau=8g^2m^3/\hbar^5$ and $\lambda=2gm/\hbar^2$ after omitting primes in $t'$, $x'$ and $\psi'$ yields directly the dimensionless GPE written in \eqref{GPE}. In terms of $\lambda$ and $\tau$,  the unit of mass is  $M=\hbar \tau/\lambda^2$.

To retrieve the dimensional predictions of Eq.~(\ref{Eq:DirectDimensional},\ref{Eq:InverseDimensional}), we first remark that $n_k$ is dimensionless and the fluxes have dimensions $[P_0]=M T^{-3}L^{-1}$ and $[Q_0]=T^{-1}L^{-3}$, and wave vectors $[k]=L^{-1}$. By reintroducing dimensional quantities in Eq.~\eqref{eq:dc} and \eqref{eq:ic} , we obtain 
\begin{subequations}
 \begin{equation}
 n_{ k} =C_{\rm d} \left(\frac{P_0}{M}\lambda\tau^3\right)^{1/3}  \left(k\lambda\right)^{-3} \ln^{-\frac{1}{3}}\left({k}/{k_{\rm f}}\right)\,,\qquad\qquad
 n_{ k} =C_{\rm i} \left(|Q_0|\lambda^3\tau\right)^{1/3} \left(k\lambda\right)^{-7/3}\,, \tag{\theequation a-b}
 \end{equation} 
 \end{subequations}
which, after replacing the values of $M$, $\lambda$ and $\tau$,  directly results in Eq.~(\ref{Eq:DirectDimensional},\ref{Eq:InverseDimensional}).

\subsection{Comparison of the experimental findings with the log-corrected KZ spectrum}

\begin{figure*}[h]
\centering
\includegraphics{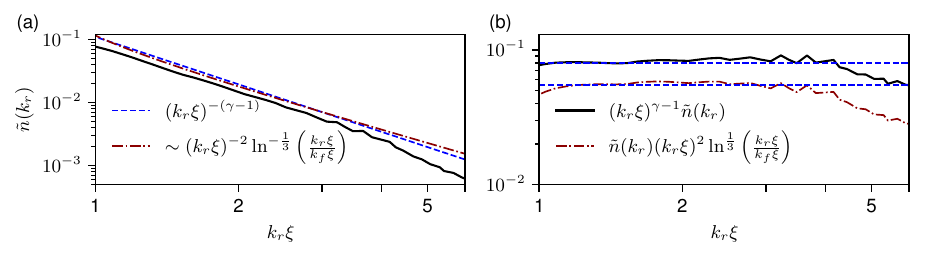}%
\caption{\label{fig:Navon}
Revision to experimental data: (a)
Comparison of experimental spectrum $\tilde{n}(k_r)=k\,n(k)$ to power law $\sim k^{-(\gamma-1)}$ with $\gamma=3.5$ and log-corrected KZ prediction (arbitrary prefactor for unknown value of flux), respectively; (b) Corresponding compensated experimental spectra.}
\end{figure*}
In Fig. \ref{fig:Navon}, we superimpose the log-corrected KZ prediction to the data extracted from Fig. 3 (a) of \cite{navon2016emergence}.
The log-corrected KZ spectrum and the  fitted power-law presented in \cite{navon2016emergence} both agree with the experimental data well for small $k_r\xi$. 
However, the experimental  inertial range reported in \cite{navon2016emergence} is limited ($1<k\xi<4$), whereas it is significantly wider in our numerical study ($1<k\xi<10$) where the superiority of the log-corrected spectrum is clearly seen.

\end{widetext}

\end{document}